\documentclass[epj]{svjour}
\usepackage{graphics}
\usepackage{psfig}
\begin{document}
\title{Stabilization of A-type Layered Antiferromagnetic Phase in LaMnO$_3$
by cooperative Jahn-Teller Deformations}
\author{M. Capone\inst{1} \ , D. Feinberg\inst{2} and M. Grilli\inst{3}
}                     
%
%
\institute{$^1$ Istituto Nazionale di Fisica della Materia and
International School for Advanced Studies (SISSA-ISAS), Via Beirut 2-4,
34013 Trieste, Italy \\
 $^2$ Laboratoire d'Etudes des Propri\'et\'es Electroniques des Solides,
Centre National de la Recherche Scientifique, associated with Universit\'e
Joseph Fourier, BP 166, 38042 Grenoble Cedex 9, France\\
$^3$ Istituto Nazionale di Fisica della Materia and Dipartimento di Fisica,
Universit\`a di Roma "La Sapienza", Piazzale Aldo Moro 2, 00185 Roma, Italy}
\date{Received: date / Revised version: date}
\abstract{It is shown that the layered antiferromagnetic order in
stoechiometric $LaMnO_3$ cannot be understood purely from electronic
interactions. On the contrary, it mainly results from strong cooperative
Jahn-Teller deformation. Those involve a compression of the $Mn-O$
octahedron
along the $c-$axis (mode $Q_3 < 0$), while alternate Jahn-Teller
deformations occur in the $ab-$plane (mode $Q_2$).
These deformations stabilize a certain type of orbital ordering. The
resulting superexchange couplings are calculated by exact diagonalization,
taking into account both $e_g$ and $t_{2g}$ orbitals. The main result is
that antiferromagnetic (ferromagnetic) coupling along the $c-$direction
($ab-$planes) can be understood only if the Jahn-Teller energy is much
larger than the superexchange couplings, which is consistent with
experiments.
This mechanism contrasts with that based on weak Jahn-Teller coupling which
instead predicts elongation along the $c-$axis ($Q_3 > 0$). The crucial
role of the deformation anisotropy $\frac{Q_2}{Q_3}$ is also emphasized.
} 
\authorrunning{M. Capone {\it et al.},}
\titlerunning{Layered AF ordering from Jahn-Teller deformations in $LaMnO_3$}
\maketitle
%
\section{Introduction}
\label{intro}

Perovskite oxides containing Mn ions have been the object of intense
interest in the recent years. In spite of being known for a very long time,
these compounds have been reconsidered in great detail owing to their
colossal magnetoresistive properties.
Starting from the "parent" phases $LaMnO_3$ (trivalent Mn) and $CaMnO_3$
(tetravalent Mn), substitutional doping has revealed an extremely rich
phase diagram. Understanding this diagram requires at least the following
ingredients :
i) strong on-site Coulomb interactions; ii) the "double exchange" mechanism
due to the interplay of $e_g$ electron itineracy and Hund's exchange with
the more localized $t_{2g}$ electron spins, which favours ferromagnetism
\cite{Zen,And,DeGen}; iii) superexchange between $t_{2g}$ electrons as well
as between $e_g$ electrons on neighbouring sites;
iv) large electron-lattice interactions, in particular due to Jahn-Teller
(JT)
effect on $Mn^{3+}$ ions \cite{Mil,Zan}.  All these elements are necessary
to understand the interplay between spin, charge and orbital ordering. The
latter
lifts the degeneracy of the $e_g$ orbitals by a cooperative Jahn-Teller
lattice deformation and leads to
tetragonal or orthorhombic deformations of the cubic structure.

Although Goodenough \cite{Good} provided long time ago a qualitative
understanding of the phase diagram of the $(La, Ca)MnO_3$ family, a full
microscopic description is still lacking.
Especially the dramatic dependence of all physical properties with very
fine tuning of the chemical composition requires a precise estimate of
the various parameters, and clear identification of the dominant mechanism
for every doping. Surprisingly enough, such an understanding is not yet
reached in the insulating antiferromagnet $LaMnO_3$, although it seems
essential before quantitatively studying the doped phases.
This phase, when fully stoechiometric, presents a layered antiferromagnetic
order, with ferromagnetic couplings (F) in two directions and
antiferromagnetic (AF) coupling in the other \cite {Wol}. The AF directions
are associated to a shortening of the $Mn-O$ bonds, leading to tetragonal
distortion, while in the F directions long and short bonds alternate,
yielding the overall orthorhombic structure.
In what follows, we shall neglect the tilting of the $Mn-O$ octahedra and
concentrate only on the $Mn-O$ bond length deformations. These can be
understood in
terms of cooperative JT effect. The corresponding lifting of $e_g$
degeneracy can be viewed as an orbital ordering, with occupied $d$ orbitals
pointing preferentially in the directions of long $Mn-O$ bonds.

Several proposals have been made to explain layered
antiferromagnetism in $LaMnO_3$.
Goodenough \cite{Good} used the picture of  "semi-covalence" where oxygen
orbitals play an essential role in overlapping empty $d$ orbitals of $Mn$
ions.  This picture, although useful for qualitative purposes,
has not received confirmation by microscopic
 calculations and does not allow to write simple enough models, for
instance based on a Hamiltonian involving only metal orbital electrons and
their basic interactions.
A microscopic description requires to identify clearly the dominant
interactions
in the problem.
In pioneering works, Kugel and Khomskii \cite{KK}, and Lacroix
\cite{Lacr} (see also earlier work by Roth \cite{Roth}), proposed that
superexchange in the presence of
$e_g$ orbital degeneracy results in ferromagnetism and orbital ordering :
Hund's rule favours in this case different orbitals on neighbouring sites
and ferromagnetic coupling. Using a simplified model with equal hopping
integrals between $e_g$ orbitals
leads to the same ordering along the three cubic lattice directions : the
resulting structure is an insulating ferromagnet, with "antiferroorbital"
ordering.
However, taking properly into account the hopping integrals between
$d_{x^2-y^2}$ (denoted $x$) and $d_{z^2}$ (denoted $z$) orbitals, Kugel and
Khomskii \cite{KK} found the correct magnetic structure.
Starting with degenerate $e_g$ orbitals, they performed a perturbative
calculation in $\frac{t}{U}$ and $\frac{J_H}{U}$ where $t$, $J_H$ and $U$
are the typical hopping integral,
the Hund coupling and the on-site repulsion
in the order. Based on the weak
electron-lattice coupling in the compound $KCuF_3$, they considered the
JT couplings as a perturbation.
As a result, orbital and magnetic ordering result from superexchange (SE)
only: Intraorbital SE dominates in the c-direction (defined as the z-axis),
leading to AF coupling, while interorbital SE dominates in the
ab-directions, yielding F coupling.
Occupied orbitals are dominantly $d_{z^2-x^2}$ and $d_{z^2-y^2}$,
therefore, as Kugel and Khomskii remark, for $Cu^{2+}$ in $KCuF_3$ (hole
orbital), JT coupling implies a shortening as experimentally
observed ($c/a < 1$).
However, for $Mn^{3+}$ ions with one electron
in the $e_g$ levels, they correctly point out
that repulsion between metal and anion orbitals, together with JT
coupling, would trigger a lengthening of the c-axis ($c/a > 1$), in
contradiction with the actual structure. In a recent work, Feiner and Oles
\cite{Feiner} reconsidered Kugel and Khomskii's model,
including both Hund's coupling betwen $e_g$ and $t_{2g}$ orbitals and the
antiferromagnetic superexchange interaction between $t_{2g}$ spins (equal
to $\frac{3}{2}$ in the ground state).
Their results confirms those of Ref.\cite{KK}: They find the correct layered
structure (which they call MOFFA), but only if the $d_{z^2}$ orbital has
lower energy than the $d_{x^2-y^2}$ one, contrarily to what happens for
electron-like orbitals (case of $LaMnO_3$).
This contradiction sets the limits of the
Kugel-Khomskii model for $LaMnO_3$.
We believe that the JT effect, on the contrary, has to be
considered from the very beginning in the model.

Essentially, the assumption that the $e_g$ degeneracy is lifted
principally by superexchange may be justified in $KCuF_3$, but is
definitely not correct in $LaMnO_3$.
In fact, this could hold only if the typical JT splitting
$\epsilon$ was much smaller than the superexchange splitting, of order
$\frac{t^2}{U}$. The latter (related to the magnetic transition
temperatures) being of the order of a few $meV$, the former is much
larger.
Although there is no precise evaluation of this quantity, this is supported
by experiment:
On the one hand, the deformations of Mn-O bonds is extremely large, more than
ten per cent, indicating that $\epsilon > k_B T$. On the other hand,
neutron scattering measurements show that
the local distorsions persist above the orthorhombic-cubic transition
at $750K$ \cite{Moussa2}.
This temperature only marks the disappearance of cooperative JT
ordering, while distorted $MnO_6$ octahedra still exist at higher
temperatures.
Photoemission \cite{dessaushen}
measurements indicate that JT splittings are
as large as a few tenth of $eV$, comparable to the electronic
hopping integrals between neighbouring sites. And optical conductivity
analysis \cite{Jung} also shows evidence of large splittings.

In these conditions, the degenerate perturbation calculation of Ref.
\cite{KK} does
not hold anymore. In a previous work we have reconsidered the problem wihin
perturbation theory \cite{us}, making the opposite assumption, i.e.
$\epsilon >> \frac{t^2}{U}$:
This means that, given the crystal deformations, due to strong cooperative
JT effect, the $e_g$ orbitals split so as to give a certain type
of orbital ordering. The orbitals stabilized at each sites are different
 from the one predicted by pure superexchange.
We have found that, depending on the values of the two JT modes
$Q_2$ and $Q_3$, different magnetic ordering could be stabilized, among
which the layered "FFA". This ordering is always stabilized if the $Q_3$
mode is positive, e.g. for dilatation in the c-direction.
But in the real case $Q_3 < 0$, FFA order is realized only if the
in-plane alternate $Q_2$ mode is sufficiently large and overcomes the
contrary effect of $Q_3$. Looking at structural numbers, one checks
that this is actually the case.
Nevertheless the system is close to the point where the FFA order becomes
unstable towards FFF. This results in the F exchange (along the ab-plane)
being larger than the AF exchange (along the c-axis).
This feature has been obtained from inelastic neutron scattering
\cite{Moussa1}, and it cannot be explained by the Kugel-Khomskii model,
which obtains on the contrary that the F superexchange is of order
$\frac{J_H}{U}$ times the AF one, thus much smaller.

The interplay between lattice distortions and magnetism has also been
investigated from ab-initio calculations of the electronic structures
\cite{Pick,Sol,Sawa}. All conclude with a prominent role of those
distortions to stabilize the actual magnetic order.
In particular, Solovyev et al. \cite{Sol} have found that the c-axis
exchange is antiferromagnetic only if the JT distortion is
sufficiently large. For $LaMnO_3$ with its very large
distortion they obtain the layered antiferromagnetic structure, but it is
close to the
border between FFA and FFF phases.
Very recent Monte Carlo calculations have also demonstrated the relevance of
the JT interaction in stabilizing the FFA magnetic order
\cite{HYMD}.

In the present work, we reconsider the problem, beyond any perturbation
theory, by exact diagonalizations on pairs of $Mn^{3+}$ sites.
The two $e_g$ orbitals are considered
together with the quantum $\frac{3}{2}$-spins due to the electrons
in the $t_{2g}$ levels. Our conclusions confirm the
essential role of JT deformations, especially the $Q_2$ mode, to
stabilize the layered AF order.
They also demonstrate that it is essential to include Hund's coupling with
$t_{2g}$ orbitals, and that the role of the intrinsic $t_{2g}$ AF exchange
is to slightly stabilize the FFA order with respect to the FFF one.

\section{The Model}

>From the discussion of the preceding section it is clear that
the basic physical ingredients required for a satisfactory
description of the Manganites should involve both Coulomb and
lattice (namely JT) interactions.
Accordingly we consider the following model
\begin{equation}
H = H_t+H_H+H_{UU'}+H_{J}+H_{JT} \label{model}
\end{equation}
with
\begin{eqnarray}
H_t &=&-\sum_{i{\bf a}\alpha \alpha' \sigma } t^{\bf a}_{\alpha \alpha'}
            c^\dagger_{i\alpha \sigma}c_{i+{\bf a}\alpha' \sigma}
\nonumber \\
H_H  &=&-J_H\sum_{i\alpha\sigma\sigma'} c^\dagger_{i\alpha \sigma}
        {\bf s}_{\sigma\sigma'}c_{i\alpha\sigma'} \nonumber \\
& \times&  \left[ {\bf S}_i +\sum_{\alpha'\ne \alpha \tilde{\sigma}
 \tilde{\sigma}'} c_{i\alpha'  \tilde{\sigma}}
        {\bf s}_{ \tilde{\sigma} \tilde{\sigma}'}
c_{i\alpha' \tilde{\sigma}'}  \right]
\nonumber \\
H_{UU'}  &=& U\sum_{i\alpha}\left(c^\dagger_{i\alpha \uparrow}
c_{i\alpha \uparrow}\right)
\left(c^\dagger_{i\alpha \downarrow}c_{i\alpha \downarrow}\right)
\nonumber \\
&+&{U'}\sum_{i\alpha\ne\alpha'\sigma\sigma'}
\left(c^\dagger_{i\alpha \sigma}c_{i\alpha\sigma}\right)
\left(c^\dagger_{i\alpha' \sigma'}c_{i\alpha' \sigma'}\right)
\nonumber \\
H_J &=& J_t\sum_{\langle ij \rangle} {\bf S}_i\cdot {\bf S}_j
\nonumber \\
H_{JT} &=&   g\sum_i \left(
c^\dagger_{i\alpha \sigma}{\bf \tau^{(3)}}_{\alpha \alpha'}
      c_{i\alpha' \sigma}Q_{3i}+
c^\dagger_{i\alpha \sigma}{\bf \tau^{(2)}}_{\alpha \alpha'}
      c_{i\alpha' \sigma}Q_{2i}\right), \nonumber
\end{eqnarray}

The first term represents the kinetic energy with the electrons
in the Manganese $3d_{x^2-y^2}$ ($\alpha=x$) or $3d_{3z^2-r^2}$
($\alpha=z$) orbitals hopping from site $i$ to the
nearest neighbor (nn)  site $i+{\bf a}$
in the ${\bf a}$ lattice direction.
Here ${\bf s}$ is the vector of Pauli matrices for spins and ${\bf \tau}$
the vector of Pauli matrices for orbital pseudospins in the ${x,z}$ basis.
Specifically, for a standard
choice of the phases for the orbital wavefunctions,
the hopping between the $x$ and the $z$ orbitals are given by
\begin{eqnarray}
t^{{\bf x},{\bf y}}_{xx}  & = & 3t; \;\;\;
t^{{\bf x},{\bf y}}_{zz}=-t;\nonumber \\
t^{\bf x}_{xz}  & = & -\sqrt{3}t ;
\;\;\;t^{\bf y}_{xz}=\sqrt{3}t \nonumber \\
t^{\bf z}_{zz} & = & -4t\;\;\; t^{\bf z}_{xx} =t^{\bf z}_{xz}=0 \; .
\label{xzhopp}
\end{eqnarray}

Together with the Hund coupling
given by $H_H$ the kinetic energy gives rise to the usual
``double-exchange'' itinerancy of the $e_g$ electrons. The (strong)
on-site Coulomb interactions,  are represented by the intraorbital
repulsion $U$ and by the interorbital $U'=U-2J_H$ term.

For simplicity here and in the following we will not distinguish
between the Hund exchange energy between electrons in the
$e_g$ and $t_{2g}$ orbitals. The antiferromagnetic superexchange
coupling between neighboring $t_{2g}$ spins is considered with
$H_J$, while the JT
interaction between the $e_g$ electrons and the (cooperative) lattice
deformation is given by the last term $H_{JT}$.
The Jahn-Teller modes are defined in terms of the short ($s$), medium ($m$)
and long ($l$) $Mn-O$ bonds by $Q_2 = \sqrt{2}(l - s)$ and $Q_3 =
\sqrt{2/3}(2m - l - s)$, the $m$ bonds lying in the $z$ direction and the
$s,l$ ones in the $x,y$ planes.

Since in the present work we will not attempt to perform any energy
minimization by including the elastic interactions due to the
lattice, we here disregard these energy terms by treating the
JT deformations ${\bf Q} = (Q_2, Q_3)$ as external fields imposed by a
lattice ordering
involving a much higher energy scale than the magnetic ones.
Therefore in the following the various magnetic couplings
will be determined in terms of {\em assigned} lattice deformations.
This viewpoint, which already guided us in the perturbative analysis of the
stability of FFA antiferromagnetism in the undoped LMO \cite{us}
is definitely justified by the experimental observation that
the JT energy splitting is much larger than all magnetic couplings.

We exactly diagonalize
the Hamiltonian in Eq. (\ref{model}) for a system of two sites
with open boundary conditions.
The two sites are located either on the same $xy$ plane or on adjacent planes
and the suitable hopping matrix elements between the various orbitals
have been considered according to expressions (\ref{xzhopp}).

The JT energy splitting  $\epsilon=g\sqrt{Q_2^2+Q_3^2}$
and the deformation anisotropy ratio  $r\equiv Q_2/Q_3$ are given external
parameters
and are fixed for any diagonalization procedure.
Once the ground state is found, the
effective exchange coupling between the total spins on the two
sites can be determined. Specifically, since the
Hamiltonian conserves the total
spin of the two-site cluster, we determine the ground
states with total spin $S_T=4, M_{S_T}=4$ and  $S_T=3, M_{S_T}=3$.
Then the magnetic coupling is given by
the energy difference $E(S_T=4,M_{S_T}=4)-E(S_T=3,M_{S_T}=3)=2J$.
Once the magnetic couplings (and particularly their sign)
along the various lattice directions are found,
the resulting magnetic phase is also determined.

\section{Results}

In order to gain insight from the physical processes underlying
the intersite magnetic couplings, we first carry out a comparison
between the results of the perturbative analysis
of the superexchange interactions
(see Ref. \cite{us}) and the exact numerical calculations.
The perturbative analysis not only was performed assuming very large
local Coulomb interactions ($U,U'$ and $J_H$ much larger
than $t$), but the additional assumption
was made  that the JT energy splitting $\epsilon$
greatly exceeds the typical superexchange energy scale of order
$t^2/U$.  In this way the ground state can safely be assumed to
be formed by just one singly occupied $e_g$ level.
 Accordingly the exact numerical calculations to be compared with
the analytic results have
been performed for $Q_3 < 0$ and $t=0.2 eV$, $U=8eV$,
$J_H=1.2eV$, $\epsilon=0.4eV$, and $J_t=0$.
\begin{figure}
\hspace{1cm}{{\psfig{figure=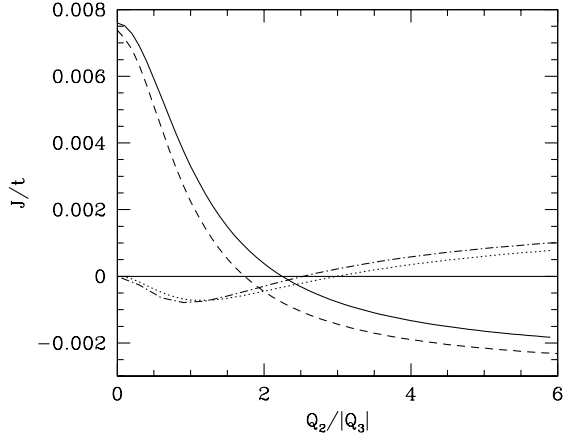,width=8cm,angle=0}}}
{\caption{
Magnetic couplings for  $Q_3 < 0$, $t=0.2 eV$, $U=8eV$,
$J_H=1.2eV$, $\epsilon=0.4eV$, and $J_t=0$
as function of the deformation anisotropy ratio
from exact diagonalization and perturbation theory. Dashed line:
$J_{xy}$ from exact diagonalization; Dot-dashed line:
$J_{z}$ from exact diagonalization; Solid line: $J_{xy}$ from perturbation
theory; Dotted line: $J_{z}$ from perturbation theory.
Negative (resp. positive) values indicate
ferro (resp. antiferro) magnetic interactions.}}
\label{Fig1}
\end{figure}
Fig. 1 reports the superexchange interactions both in
the planar and interplanar directions obtained with
both the perturbative and the exact-diagonalization analysis.
As it is apparent, the perturbative $J_{xy}$ and $J_z$
display the same
qualitative behavior as in the exact calculation. This confirms that,
at least in the $\epsilon \gg t^2/U$ limit, a substantial part of
the magnetic effective interactions is generated by the
superexchange processes due to the hopping of electrons
lying in the lower $e_g$ level on the same or
on different nearest neighbor $e_g$ orbitals. On the other
hand, the quantitative comparison indicates that the range
of stability for the FFA phase (i.e. $J_{xy}<0$ and $J_z>0$)
is modified. In fact a positive $J_z$ together with a negative
$J_{xy}$ are obtained in the exact calculation on a somewhat larger
range of lattice deformation anisotropies
$(Q_2/|Q_3|\ge 2.5)$. In order to establish
a tighter connection
between the experimentally determined $J$'s and the observed
deformations, and to investigate the role of the various
interactions in the model, a more systematic
analysis is required.
Assuming the JT interaction to be relevant in
stabilizing the FFA phase, we investigate the behavior
of the exchange constants $J_{xy}$ (denoted "intraplane") and $J_z$
(denoted "interplane") in terms of $\epsilon$ and the deformation
ratio $r$.

Figs. 2 and 3 report $J_{xy}$ and $J_z$ as functions of
$|r|$ for the $Q_3<0$ case (the one relevant for LMO) at a large
($\epsilon \sim 2t$) and at a small ($\epsilon <0.1t$)
value of the JT splitting respectively. Different values of
the Hund coupling $J_H$ are considered.
\begin{figure}
\hspace{1cm}{{\psfig{figure=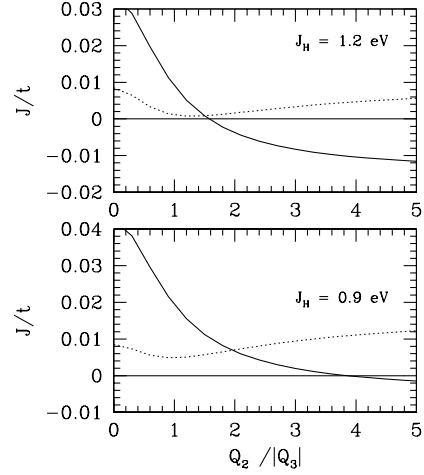,width=8cm,angle=0}}}
{\caption{Magnetic couplings $J_{xy}$ (solid line) and $J_z$ (dotted line)
 vs. the deformation ratio $r=Q_2/|Q_3|$
for negative values of $Q_3$,
for  $t=0.14eV$, $U=6eV$, $\epsilon=0.3eV$, $J_t=2.1meV$ and
$J_H=1.2eV$ ($J_H=0.9eV$) in the upper (lower) panel.
}}
\label{Fig2}
\end{figure}
\begin{figure}
\hspace{1cm}{{\psfig{figure=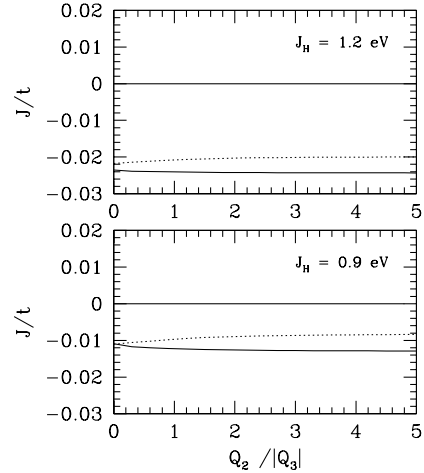,width=8cm,angle=0}}}
{\caption{Magnetic couplings (see Fig. 2) vs. the deformation ratio
$r=Q_2/|Q_3|$
for negative values of $Q_3$,
for  $t=0.14eV$, $U=6eV$, $\epsilon=0.01eV$, $J_t=2.1meV$ and
$J_H=1.2eV$ ($J_H=0.9eV$) in the upper (lower) panel.
}}
\label{Fig3}
\end{figure}
One can first observe
from Fig. 2 that in the large-$\epsilon$ case the
increase of the Hund coupling shifts downwards
both the intraplane and the interplane magnetic couplings.
This outcome can be rationalized in terms of perturbatively
generated superexchange processes  providing
AF effective couplings of the form
\begin{equation}
J^{AF}_{xy,z} \approx \frac{A_{xy,z}}{U+(3/2)J_H}+
 \frac{B_{xy,z}}{U+\epsilon}
\label{SEAF}
\end{equation}
competing with the generated F interaction
\begin{equation}
J^{F}_{xy,z} \approx -\frac{C_{xy,z}}{U+\epsilon-(5/2)J_H}.
\label{SEF}
\end{equation}
The numerical coefficients $A,B$, and $C$ stem from the different
hopping matrix elements between the different orbitals in the different
directions \cite{us}. Specifically, while the $A$'s are related to the
hopping processes between two nearest-neighbour lower-lying $e_g$ orbitals,
the $B$ and $C$ coefficients are due to hoppings between one
low-lying and one higher JT-split orbitals (this is why the corresponding
denominators involve $\epsilon$). The $A,B$ and $C$ coefficients
are independent from the Coulomb interactions,
which only determine the energies of the virtual intermediate states
in the superexchange processes. The above schematic expressions
clearly show that, when $J_H$ is increased for a fixed
$\epsilon \gg t$, the F spin configuration becomes more
favourable, since the F coupling become stronger, while the
AF interaction weakens. We remark that purely electronic models such as in
Refs.\cite{KK,Feiner} make use of degenerate perturbation theory. Then the
orbital splitting is of order of the exchange couplings $J$ and therefore
those models become invalid if $\epsilon > J$, which is the case in
$LaMnO_3$.
More seriously, the orbital order resulting
from purely electronic interactions is at odds with that obtained from the
actual Jahn-Teller distortions, showing that those distortions do not
result from an orbital ordering of electronic origin, but are on the
contrary the mere source of orbital ordering.

Another quite generic effect, which can be interpreted in terms
of perturbatively generated superexchange processes is the
tendency of $J_z$ to acquire a F (or at least a less AF) character
at low values of $Q_2/|Q_3|$ (this can also be accompanied by
an upturn of $J_z$ for $|r|$ tending to zero).  This occurs because
for $Q_3<0$, the lowest $e_g$ level progressively loses its
$3d_{3z^2-r^2}$ component: By schematically
writing the lower and the upper $e_g$ states as $|a\rangle \propto
|x\rangle + \eta |z\rangle$ and $|b\rangle \propto
- \eta|x\rangle + |z\rangle$ respectively,   $\eta$ vanishes
with $|r|\to 0$. Now, the superexchange along $z$ is driven by
the interplane hopping, which is only allowed between  $3d_{3z^2-r^2}$
orbitals. Furthermore one can see \cite{us} that the ferromagnetic
superexchange arises from $|a\rangle \to |b\rangle$
hoppings, which are of order  $\eta$, while the antiferromagnetic
coupling is mostly generated from intraorbital $|a\rangle \to |a\rangle$
hopping (the $A$ term in Eq. (\ref{SEAF}).
Since this latter is of order $\eta^2$, it is quite
natural that in the low-$|r|$ region, as $\eta$ decreases,
the superexchange along $z$ is ferromagnetic
and vanishes with $\eta$. This ferromagnetic tendency is, however,
contrasted (and actually overcome in Figs. 2 and 3)
by the independent AF superexchange $J_t$ between
the $t_{2g}$ spins, which becomes relatively more important.
Of course, when $|r|$ increases, the $\eta^2$ terms in the hopping
become relevant, the intraorbital $|a\rangle \to |a\rangle$
hopping starts to dominate and $J_z$ eventually becomes (more)
positive (i.e. AF).

As far as the superexchange along the planes is concerned, at small $|r|$
this is instead dominated by the large hopping between
$3d_{x^2-y^2}$ orbitals, which favor the $|a\rangle \to |a\rangle$
hopping and, consequently produces an AF magnetic coupling.
On the contrary, for large $|r|$, orbital ordering implies that the main
superexchange contribution comes from hopping between different orbitals,
thus favouring ferromagnetism \cite{KK}.

All the above arguments are obviously only valid as long as the
conditions for the perturbation theory nearly hold.
On the other hand, the simple perturbative
approach between non-degenerate states breaks down when
$\epsilon \approx t^2/U$ as in Fig. 3 and the interpretation of the
results is not so transparent. However, the effect of $J_H$
favoring ferromagnetism is still present.

An important difference between the results in Figs. 2 and 3 is that
the FFA phase is generically obtained in a broad range of parameters
when $\epsilon \gg t$.
In particular, for rather realistic values of $J_H\sim 5t \approx 1$eV
the deformation ratios required to generate negative (i.e. F) couplings
in the $xy$ planes and positive ones in the z direction are quite reasonable
$|r| \sim 2-3$. The same does not hold in the case of small JT splitting,
where $J_{xy}$ and $J_z$ have the same sign (FFF).
Therefore a first result is that
a sizable $\epsilon$ is needed in order to obtain both the FFA phase
and reasonable lattice distortion ratios $Q_2/Q_3$.

This result is also confirmed by the calculation of $J_{xy}$ and $J_z$
as a function of $\epsilon$,
at a fixed value of the deformation anisotropy ratio $r$.
Figs. 4 and 5 report the values of $J_{xy}$ and $J_z$
for $r=3$ and $r=-3$ respectively.
\begin{figure}
\hspace{1cm}{{\psfig{figure=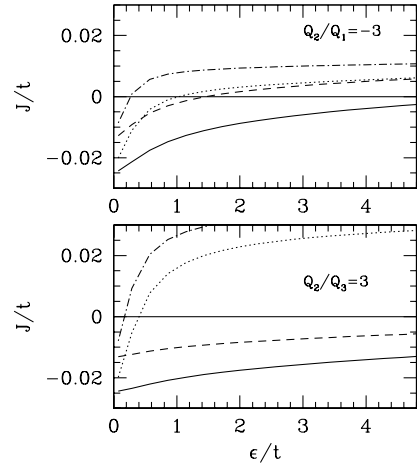,width=8cm,angle=0}}}
{\caption{Magnetic couplings vs. the JT splitting energy $\epsilon$
for  $t=0.14eV$, $U=6eV$, $J_t=2.1meV$ for $Q_2/Q_3=-3$ (upper
panel) and $Q_2/Q_3=3$ (lower panel). Solid line: $J_{xy}$ for
$J_H=1.2eV$; Dotted line:  $J_{z}$ for $J_H=1.2eV$;
Dashed line: $J_{xy}$ for $J_H=0.9eV$; Dot-dashed line:
$J_{xy}$ for $J_H=0.9eV$.
}}
\label{Fig4}
\end{figure}
While the positive $r$ case is
generic for perovskite materials with the lattice elongated in the
$z$ direction ($c/a>1$), the latter choice is more pertinent to the
case of the undoped LMO, where $c/a<1$.
As already discussed by Kugel and Khomskii \cite{KK}
for a different model and as confirmed by the
perturbative analysis of Ref. \cite{us}, the JT deformation
and the superexchange interactions cooperate when $Q_3>0$ like in
KCuF$_3$ so that it is not surprising that for all values of $J_H$
the FFA is realized over a much broader range of $\epsilon$.
On the other hand, for $Q_3<0$, Fig. 4 shows that the conditions
for a FFA phase, $J_{xy}<0$ and $J_z>0$, are only realized for
a smaller range of $\epsilon$ values.
In particular a sizeable minimum value of $\epsilon$
is required to have an AF coupling along $z$, while exceedingly
large values of $\epsilon$ (of order $J_H$)
produce an AF coupling also along the planes.
Both the minimum and the maximum values of $\epsilon$ for obtaining
the FFA phase increase upon increasing $J_H$. However, the
maximum value of $\epsilon$ increases more rapidly and the overall effect
is that, increasing  $J_H$, the available range in $\epsilon$
to obtain an FFA phase is enlarged.
Again the behavior displayed in the exact calculations reported in
Fig. 4 can easily be interpreted
in terms of the perturbative superexchange processes schematically
represented in Eqs. (\ref{SEAF}) and (\ref{SEF}).
First of all these expressions at once account for the
increasing behavior of the couplings upon increasing $\epsilon$:
While only the interorbital part of $J_{AF}$ (the contribution
proportional to $B$) decreases upon increasing $\epsilon$, the
whole ferromagnetic part in Eq. (\ref{SEF}) is suppressed
when $\epsilon$ grows, so that the total coupling, although
ferromagnetic at small JT energy splitting, eventually
vanishes and becomes positive.

Moreover it turns out that, for $|r| > 2-3$
the hoppings generate smaller $A,B,C$ coefficients in the $z$
direction. This accounts for the more rapid rise of $J_z$
when $\epsilon$ is increased. Finally, along the same line of the
discussion of Fig. 2,  one can easily observe
that an increasing $J_H$ strenghtens the ferromagnetic component
and weakens the antiferromagnetic one, thus rationalizing the generic
tendency of all curves to be shifted downwards when $J_H$ grows.

Besides the above specific findings, the occurrence of
the various magnetic phases can be cast in a phase diagram
at zero temperature illustrating the stability region
of these phases in terms
of the JT energy splitting and the deformation ratio.
In the light of their richer complexity and of the present interest
in the Manganites, we here consider in greater detail the case of $Q_3<0$
of relevance for the undoped LMO, while the $Q_3>0$ case is only
described in the inset of Fig. 5.
Figs. 5 and 6 report the phase diagram for
two different values of the Hund coupling.
\begin{figure}
{{\psfig{figure=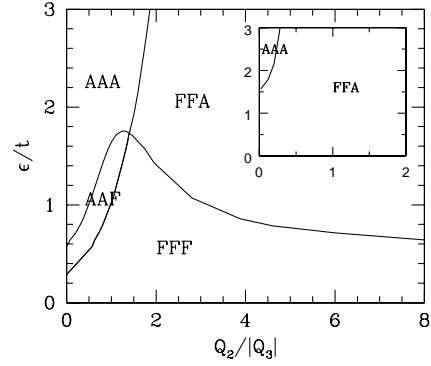,width=8cm,angle=0}}}
{\caption{Zero-temperature phase diagram for $t=0.14eV$, $U=6eV$, $J_t=2.1meV$
and $J_H=1.2eV$.}}
\label{Fig5}
\end{figure}
\begin{figure}
{{\psfig{figure=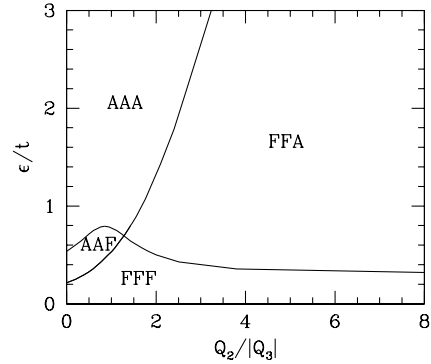,width=8cm,angle=0}}}
{\caption{Zero-temperature phase diagram for $t=0.14eV$, $U=6eV$, $J_t=2.1meV$
and $J_H=1.0eV$.}}
\label{Fig6}
\end{figure}
Both phase diagrams display the same qualitative features. In particular,
at moderate and large values of $\epsilon$ a N\'eel AAA phase
is found for weak planar distortions (small $r$).
As seen in the discussion of Fig. 2, in the very-small-$r$ region,
$J_{xy}$ is naturally positive, while
the superexchange between $e_g$ levels along $z$, although
ferromagnetic, is small so that
the direct superexchange between $t_{2g}$ spins may easily dominate and
gives rise to the AAA phase (see Figs. 2 and 3).
As it can be also be seen from Fig. 1, it can be checked
that the AAA phase is replaced by the so-called $C$-like antiferromagnetic
AAF phase
in the $J_t=0$ case. At small-to-intermediate values of $\epsilon$,
a progressive increase of $|r|$ drives the
system towards the phase
AAF. In this phase $J_{xy}$ keeps its AF character, while
the negative superexchange between $e_g$ levels along $z$
is small, but no longer is overcome by $J_t$.
At larger values of $\epsilon$ the AAF phase is not present, but
the intimate nature of the AAA phase changes upon increasing $|r|$.
In particular while at low $|r|$ the AF along $z$
is determined by $J_t$, at larger $|r|$, the superexchange between
$e_g$ levels along $z$ is itself AF and therefore
the $t_{2g}$ superexchange contributes, but it is not strictly
necessary to the AF coupling along $z$. On the other hand,
a further increase
of $|r|$ promotes a F coupling along the planes and
leads to the A-type antiferromagnetism FFA experimentally
observed in undoped LMO.

At small values of the JT splitting,
the phase diagram is prominently occupied by a FFF phase.
In this latter regard, from the comparison of Figs. 5 and 6,
the  important observation can be done that the FFF phase
at low and moderate $\epsilon$'s is greatly stabilized by the
increase of the Hund coupling $J_H$, as previously expected.

Within the present exact numerical treatment of the model
in Eq. (\ref{model}) it is also possible to attempt at
``precise'' estimates of $J_{xy}$ and of $J_z$.
 As an example, we report here  a realistic
sets of parameters (among many others) providing the values
$J_{xy}=-0.83$ meV and $J_z=0.58$ meV
experimentally observed with inelastic neutron scattering \cite{Moussa1}.
Assuming $Q_2/|Q_3|=3.2$, a value largely confirmed
 by many groups \cite{Moussa2}, we take $t=0.124eV$, $U=5.81eV$,
$J_H=1.2eV$, $J_t=2.1$ meV, and $\epsilon=0.325eV$.
The quite reasonable values
of the model parameters needed to reproduce the
measured magnetic couplings is an indirect test of the
validity of the considered model. We emphasize that the "anomalous" trend
$|J_{xy}| > |J_z|$ is correctly reproduced, and that our fit is relatively
flexible concerning parameters $U$, $J_H$ or $t$, provided $\epsilon$ is
large enough.

\section{Conclusions}

In this paper we presented the results of calculations based on the
exact diagonalization
of a model aiming to describe the stoechiometric LaMnO$_3$. The model
includes strong local Coulomb interactions as well as a JT coupling
between the electrons and the $Q_2$ and $Q_3$ lattice deformations.

Despite the smallness of our cluster, we believe that our
determination of the magnetic couplings not only is
qualitatively, but also quantitatively significant. This is
so because, in the presently considered undoped LMO,
the coherent charge mobility is negligible due to the large on-site Coulomb
repulsions and to the substantial JT deformations. As a consequence
the magnetic interactions do not arise, e.g., from Fermi surface instabilities
or other collective effects, but are rather determined by short-distance
(incoherent) processes.

One first relevant result is that, when the $Mn-O$ octahedron is compressed
along $z$, a FFA phase is only obtained for a sizable (staggered) $Q_2$
deformation of the planar unit cell. This finding agrees with
the {\it ab initio} calculations of Ref. \cite{Sol}.

Our analysis also points out the relevant role played by the
Hund coupling, which generically emphasizes the ferromagnetic
component of the superexchange processes. Quite relevant turns
out to be also the Hund coupling between the $e_g$ electrons
and the $t_{2g}$ spins. In this latter regard, we explicitely
checked that, keeping $J_H$ finite between the $e_g$ electrons,
but decoupling them from the $t_{2g}$ spins no longer gives rise
to the FFF phase at low values of the JT splitting (cf. the
phase diagrams in Figs. 6 and 7). Instead at $\epsilon \sim 0$
a FFA phase is found in agreement with the results of Ref. \cite{KK}
for a model, which only considered $e_g$ electrons and no JT splitting.
This indicates that the determination of
the stable phase (at least) at small values of the JT
energy must take in due account the Hund coupling thereby including
the $t_{2g}$ levels. Secondly a quantitative
determination of the stability region for the FFA phase and of the
value of the magnetic couplings is subordinate to the
consideration of the $J_H$ term.

Our work shares with Ref. \cite{HYMD} the generic result that
JT distortions strongly affect the magnetic structure. Nevertheless
it is worth pointing out some differences. In a certain respect our work
is less ambitious in so far it does not attempt to determine the
JT distortions, but it rather imposes them as external parameters
of the calculation. Actually we do not believe that such deformations
can be easily determined by
microscopic models, which should incorporate complex effects such as long-range
Coulomb interactions, cation and anion size and tilts of the
$MnO_6$ octaedra. On the other hand realistic deformations
as obtained from experiments can easily be imposed and the consequent
local electronic structure can be determined exactly: orbital ordering
results essentially from cooperative Jahn-Teller deformations.

Moreover, and quite importantly for a
quantitative determination of the magnetic coupling and of the
stability of the magnetic phases, we here also take into account the
electronic Coulomb repulsion. This interaction is perforce larger than the
JT interaction and contributes
to its insulating behavior as well as to the numerical values of the
exchange couplings.

Finally we showed that using reasonable parameters the experimental
values of the magnetic couplings can easily be reproduced. Of course
precise estimates depend on the knowledge
of the various couplings entering the model, which are not always
available neither from experiments nor from reliable first principle
calculations. However
calculating the magnetic couplings for  various
parameters and matching the
numerical results with the experimentally obtained values
provides useful connections between the involved
parameters and set limits to the poorly known
physical quantities.


%
%
%
%
%

\end{document}